\documentclass[conference]{IEEEtran}

\usepackage{cite}
\usepackage{graphicx}
\usepackage{comment}
\usepackage{balance}
\usepackage{acronym}
\usepackage{amsmath}
\usepackage{tabularx}
\usepackage[usenames,dvipsnames]{xcolor}
\usepackage{fixltx2e}
\usepackage[hang]{subfigure}
\usepackage{booktabs}
\ifCLASSINFOpdf

\else

\fi

\title{Single microphone speaker extraction using unified time-frequency Siamese-Unet}

\author{\IEEEauthorblockN{Aviad Eisenberg}
\IEEEauthorblockA{Bar-Ilan University, OriginAI}
\and
\IEEEauthorblockN{Sharon Gannot}
\IEEEauthorblockA{Bar-Ilan University}
\and
\IEEEauthorblockN{ Shlomo E. Chazan}
\IEEEauthorblockA{OriginAI, Bar-Ilan University}}

\begin{document}

\newcommand{\noteb}[1]{\textcolor{blue}{#1}}
\newcommand{\noter}[1]{\textcolor{red}{#1}}
\newcommand{\noteg}[1]{\textcolor{green}{#1}}
\newcommand{\noteo}[1]{\textcolor{olive}{#1}}
\newcommand{\notem}[1]{\textcolor{magenta}{#1}}
\newcommand{\norm}[1]{\left\lVert#1\right\rVert}

\def\y{y(t,k)}
\def\sd{s_d(t,k)}
\def\hd{h_d(t,k)}
\def\rd{r_d(t,k)}
\def\si{s_i(t,k)}
\def\hi{h_i(t,k)}
\def\hr{h_r(t,k)}
\acrodef{STFT}{short-time Fourier transform}
\acrodef{ISTFT}{inverse short-time Fourier transform}
\acrodef{BSS}{Blind Source Separation}
\acrodef{DOA}{Direction of Arrival}
\acrodef{DC}{Deep Clustering}
\acrodef{DPRNN}{dual-path recurrent neural network}
\acrodef{TF}{Time-Frequency}
\acrodef{TCN}{Temporal Convolutional Network}
\acrodef{ATF}{Acoustic Transfer Function}
\acrodef{SI-SDR}{Scale-Invariant Signal-to-Distortion Ratio}
\acrodef{MSE}{Mean Square Error}
\acrodef{DFT}{Discrete Fourier Transform }
\acrodef{SIR}{signal-to-interference ratio}
\acrodef{OVA}{overlap-and-add}
\acrodef{SDR}{signal-to-distortion ratio}
\acrodef{BLSTM}{Bidirectional Long Short-Term Memory}
\acrodef{SOTA}{state-of-the-art}
\acrodef{RI}{Real-Imaginary}
\acrodef{RIR}{Room Impulse Response}
\acrodef{SNR}{signal-to-noise ratio}
\acrodef{RNN}{Recurrent Neural Networks}

\maketitle

\begin{abstract}
In this paper\footnote{This project has received funding from the European Union’s Horizon 2020 Research and Innovation Programme under Grant Agreement No. 871245.} we present a unified time-frequency method for speaker extraction in clean and noisy conditions. Given a mixed signal, along with a reference signal, the common approaches  for extracting the desired speaker are either applied in the time-domain or in the frequency-domain. In our approach, we propose a Siamese-Unet architecture that uses both representations. The Siamese encoders are applied in the frequency-domain to infer the embedding of the noisy and reference spectra, respectively. The concatenated representations are then fed into the decoder to estimate the real and imaginary components of the desired speaker, which are then inverse-transformed to the time-domain. The model is trained with the \ac{SI-SDR} loss to exploit the time-domain information. The time-domain loss is also regularized  with frequency-domain loss to preserve the speech patterns. Experimental results demonstrate that the unified approach is not only very easy to train, but also provides superior results as compared with \ac{SOTA} \ac{BSS} methods, as well as commonly used speaker extraction approach. 
\end{abstract}

\IEEEpeerreviewmaketitle

\section{Introduction}
Extracting a desired speaker from a mixture of overlapping speakers is a challenging task that is usually solved using microphone array processing~\cite{gannot2017consolidated}. With a single microphone, no spatial information is available thus making the task even more challenging. In this paper, we address the single-microphone speaker extraction task and focus on the extraction of a single participant from a mixed signal, given a pre-recorded sample of the speaker to be extracted.

In recent years a significant progress has been achieved in the field of speaker separation.
The Conv-Tasnet algorithm \cite{luo2019conv} is applied in the time-domain. Self-learned representations of the signal are inferred using 1-D conventional layers. The model estimates a mask for each speaker, which is then applied in the learned-representations domain for the separation task. The gist of this algorithm is the use of the \ac{SI-SDR} loss function \cite{le2019sdr}, which is designed to exploit the time-domain information. The authors show that by being independent of the traditional hand-crafted features, the an improved performance is obtained.
The \ac{DPRNN} algorithm, also applied in the time-domain, was presented in \cite{luo2020dual}. 
The mixing signal is split into overlapped chunks and processed using intra-chunk and inter-chunk \acp{RNN}. The performance of the algorithm was evaluated with the WSJ0-2mix database. 
In \cite{chazan2021single} an algorithm that can successfully process mixtures with larger number of speakers is presented. Moreover, it can also work  in noisy and reverberant scenarios. The algorithm comprises a multi-head architecture, jointly trained with a gate. Each head is responsible to separate different number of speakers. The gate, which classifies the number of speakers in the mixture, determines which of the heads should be applied to the mixture.

While these algorithms demonstrate promising results, they  suffer from excess computational complexity, due to the need to train the feature extraction stage, rather than to utilize the traditional \ac{TF} representation. Additionally, the permutation problem \cite{kolbaek2017multitalker} must be taken into consideration during training. Finally, it is reasonable to assume that additional information regarding the specific characteristics of the desired speaker may be beneficial in accomplishing the task of its extraction.   

Recently, different architectures were proposed to extract the desired speaker given a reference signal. They can be roughly split into \ac{TF}-domain methods and time-domain methods.
In the \ac{TF}-domain, most models are using masking operation on the mixed signal \cite{vzmolikova2019speakerbeam , wang2018VoiceFilter , he2020speakerfilter , Li2020 , Wang2018 , xiao2019single}.
In \cite{wang2018VoiceFilter} a pre-trained d-vector \cite{wan2018generalized} is utilized as an embedding of the reference signal. A mask is then estimated given the mixed signal and the reference embedding vector. Finally, the noisy \ac{STFT} is multiplied with the mask to extract the desired speaker. Note that the phase of the mixed signal is not processed, and only the spectrogram of the desired speaker is extracted.
The permutation problem is not an issue in the problem of speaker extraction, as a prior information on the desired signal is available. Yet, since applied in the \ac{TF} domain, these methods use the \ac{MSE} loss as their training objective rather than the time-domain \ac{SI-SDR} loss, which is perceptually more meaningful. 


These drawbacks, namely the use of the \ac{MSE} loss and of the noisy phase,  have led to a series of models applied directly in the time-domain \cite{xu2019time , xu2020spex , zhang2020x , delcroix2020improving , ge2020spex+ , han2021attention,deng2020robust}. Inspired by the time-domain \ac{BSS} algorithms, the architecture of these methods comprises an encoder block, a separation block (usually based on a proven \ac{BSS} architecture) and a decoder block to implement the inverse-transform of the desired signal back to the time-domain. These methods, similar to the time-domain \ac{BSS} algorithms, also utilize  the \ac{SI-SDR} loss function \cite{le2019sdr} as their objective. Unfortunately, similar to the time-domain \ac{BSS} methods, these extraction methods suffer from two main drawbacks:  1) thay are not easy to implement, and 2) they ignore the specific \ac{TF} patterns of the speech signal.

In this paper, we propose a Siamese-Unet architecture that uses
both representations, the \ac{TF} features as input and output features to the network along with the time-domain representation for computing the \ac{SI-SDR} loss. Our model is constructed with a two-head encoder, one for the reference signal and the other for the mixed signal. The estimated embeddings are then concatenated as an input to the decoder, to extract the desired speaker. 
The \ac{RI} components of the \ac{STFT} are utilized as the input while the waveform is utilized as the target of the network. The \ac{RI} components are inverse-transformed to the time-domain and the \ac{SI-SDR} loss is used to train the model. A comprehensive simulation study using common databases demonstrates the benefits of the proposed scheme.

\section{Problem Formulation}

Let $x(t)$ be a mixture of $I$ concurrent speakers captured by a single microphone:
\begin{equation}
    x(t)=\sum^I_{i=1}  \{s_i\ast {h}_i\}(t) +n(t) \quad {t=0,1,\ldots,T-1}
    \label{eq:mix_time}
\end{equation}
where $s_i(t)$ represents  the signal of the $i$-th speaker,  ${h}_i(t)$ represents the \ac{RIR} between the $i$-th speaker and the microphone, and $n(t)$ represents the additive noise. Note that in a noiseless and anechoic enclosure, ${h}_i(t)= \delta(t),\,i=1,\dots,I$ and $n(t)=0$.
In the \ac{STFT} domain \eqref{eq:mix_time} can be formulated as, 
\begin{equation}
x(l,k) =  \sum^I_{i=1} s_i(l,k) \cdot h_i(l,k) +n(l,k)
\label{eq:mix_STFT}
\end{equation}
where $l \in \{0,\ldots, L-1\}$ and $k \in \{0,\ldots, K-1\}$ are the time-frame and the frequency-bin (TF) indexes, respectively. The terms $L$ and $K$ represent the total number of time-frames and frequency bands, respectively. 

For simplicity, we address in this paper the $I=2$ case and denote the desired speaker as $ s_d(l,k) $, the reference signal as $  s_r(l,k)$ and the interference speaker as  $  s_i(l,k)$.
The output of the proposed algorithm is $\hat{s}_d(l,k)$, an estimate of ${s}_d(l,k)$ given the mixed signal \eqref{eq:mix_STFT} and a \emph{reverberant} reference signal $s_r(l,k) \cdot h_r(l,k)$.

\section{Proposed Model}
In this section we introduce the proposed novel Siamese-Unet architecture for desired speaker extraction, given a reference recording. 

\begin{figure*}[h]
\centering
    \includegraphics[width=17cm, height=5.8cm]{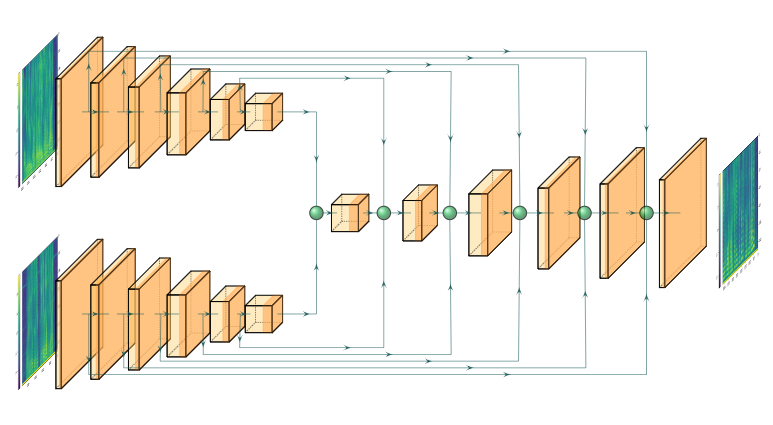}
    \caption{The proposed architecture. The green circles stands for the concatenation operation. To calculate the \ac{SI-SDR} loss, \ac{ISTFT} is applied to the model's output. The inputs and outputs features of the network are the \ac{RI} components of the mixture, reference and estimated signals, respectively. }
    \label{fig:diagram}
\end{figure*}

\subsection{Architecture}
The proposed Siamese-Unet model comprises a two-head encoder and a decoder. Skip connections are concatenated between the layers of the encoders and the decoder sections. The proposed architecture is summarized in Fig.~\ref{fig:diagram}. 

In our approach, the reference signal and the mixed signal are first projected to the same latent space by the two-head encoder. 
The encoder's architecture is constructed with seven convolution layers, each layer followed by a two-dimensional batch normalization and a `Relu' function. The decoder architecture is similar to the encoder architecture, but instead of the convolution layers, transpose-convolution layers are applied.
Denote CBR\textsubscript{$i,o$} and TCBR\textsubscript{$i,o$} as the Convolution-BatchNormalization-Relu and the Transpose-Convolution-BatchNormalization-Relu layers, respectively, where $i$ and $o$ are the number of the input and output channels, respectively. The size of the filters in each layer is set to 4, the stride size is set to 2 and the padding value is set to 1. 

The encoder path is given by:
CBR\textsubscript{64,128} $\rightarrow$ CBR\textsubscript{128,256} $\rightarrow$ CBR\textsubscript{256,512} $\rightarrow$ CBR\textsubscript{512,512} $\rightarrow$ CBR\textsubscript{512,512} $\rightarrow$ CBR\textsubscript{512,512} $\rightarrow$ CBR\textsubscript{512,512} 
and the decoder path is given by:
TCBR\textsubscript{1024,512} $\rightarrow$ TCBR\textsubscript{1536,512} $\rightarrow$ TCBR\textsubscript{1536,512} $\rightarrow$ TCBR\textsubscript{1536,256} $\rightarrow$ TCBR\textsubscript{768,128} $\rightarrow$ TCBR\textsubscript{384,64} $\rightarrow$ TCBR\textsubscript{192,2} 

Finally, an additional convolution-layer is applied to obtain the desired signal estimate.



Different alternatives for integrating the information from the reference signal are described in \cite{vzmolikova2019speakerbeam}. 
Two comments regarding the implementation are in place: 1) when using the skip connections in the U-net architecture, concatenating the encoder layers and the decoder layers, rather than multiplying them, yields better results, and 2)
concatenating all intermediate layers of the reference encoder using skip connections (rather than only the bottleneck layer) in parallel to the skip connections of the mixture encoder, improves separation performance. While the majority of \ac{STFT}-domain algorithms are  applying a mask to the mixed signal, our proposed network is directly trained to estimate the \ac{TF} representation of the target source. The structure of the network is depicted in Fig.~\ref{fig:diagram}. 

\subsection{Features}

As mentioned above, the gist of this paper is the utilization of both the \ac{TF} and the time-domain information. Most of the  approaches applied in the \ac{STFT} domain use the noisy phase for calculating the inverse-transform back to the time domain, since estimating the phase is a cumbersome task. Unfortunately, the performance of such approaches is limited even if the spectrogram is perfectly estimated, especially in reverberant environment. 
Instead, we propose to use the \ac{RI} components as both the input features and the model's target.  In this way, we circumvent the inaccuracies that result from applying the inverse-\ac{STFT} with the noisy phase.



\subsection{Objectives}
To train the proposed Siamese-Unet for the extraction task, the time-domain \ac{SI-SDR} loss function, which was found to be most appropriate for \ac{BSS} tasks, is used. The loss is formulated as,
\begin{equation}
\text{SI-SDR}\left( s,\hat{s} \right)= 10 \log_{10} \left( \frac{\norm{\frac{\langle {\hat{s},s} \rangle}{\langle {s,s} \rangle} s}^2}{\norm{\frac{\langle {\hat{s},s} \rangle}{\langle {s,s} \rangle} s-\hat{s}}^2} \right)
\end{equation}
where $\langle {,} \rangle$ is the inner product, $s$ is the target speaker in the time domain and $\hat{s}$ is the estimated speaker. 

To further improve the training, we used, for each training sample, the same mixture and swapped the desired and  interference signals. The corresponding reference signal was used for each of the extracted sources. The two losses are then averaged, 
\begin{equation}
L_{\text{SISDR}} =0.5\cdot \left[ \text{SISDR}\left( s_1,\hat{s}_1 \right) + \text{SISDR}\left( s_2,\hat{s}_2 \right) \right].
\label{eq:loss_sisdr}
\end{equation}
We also add the \ac{MSE} loss as a regularization term to the \ac{SI-SDR} loss, considering the \ac{RI} features,
\begin{equation}
    L_{\text{MSE}} =0.5\cdot \left[ \text{MSE}(\text{RI}_1,\widehat{\text{RI}_1}) + \text{MSE}(\text{RI}_2,\widehat{\text{RI}_2}) \right].
    \label{eq:loss_mse}
\end{equation}
Our final training loss is a weighted sum of the main loss and its regularization term:  
\begin{equation}
    L = \beta_{\text{SISDR}}\cdot L_{\text{SISDR}} + \beta_{\text{MSE}} \cdot L_{\text{MSE}}
\end{equation}
with $\beta_{\text{SISDR}}+\beta_{\text{MSE}}=1$.

To the best of our knowledge, this is the first work to combine both time and time-frequency representations in the training of the network. By doing so, we preserve the \ac{TF} patterns of the speech signal, while still optimizing the perceptually meaningful \ac{SI-SDR} loss. 
As a byproduct, we found that this method is easier to implement and that its training time is faster than the respective training time of algorithms with only time-domain loss. 

\section{Experimental study}
In this section, we describe the experimental setup, the train and test datasets, and the obtained results. The performance of the proposed algorithm and the competing methods is reported for both clean conditions and for mild noise and reverberation conditions.

\subsection{Datasets}

To train our model, we constructed a dataset of mixed signals. Each sample in the dataset consists of a simulated mixture, two reference signals and two clean signals, used as targets to the model.

\noindent\textbf{Dataset of  mixed signals in clean conditions:} The clean speech signals were randomly drawn from the LibriSpeech corpus \cite{panayotov2015librispeech} and the WSJ0 corpus \cite{paul1992design}.  
Each corpus was randomly split, with $80\%$ of the speakers taken for the training dataset, $10\%$ for validation dataset and $10\%$ for test.

The signals in the training phase were randomly truncated to a duration of 2-8 seconds. The duration is kept fixed for each batch and may vary between batches. 
In the test phase, the mixed signals were not truncated. If the reference sentence is shorter than the mixed signal, it is repeated until it fits the duration of the mixture. 
A total of 50,000 training samples, 10,000 validation samples and 3000 test samples were simulated. The gender of the speakers was uniformly selected. Finally, the signals are summed up to generate the mixing signal.

\noindent\textbf{Dataset of  mixed signals in mild noise and reverberation conditions:} We also constructed a noisy and reverberant dataset. Signals were randomly drawn from the LibriSpeech and WSJ0, following a similar procedure to the construction of the clean dataset. Each signal was convolved  with a simulated \ac{RIR}  using the RIR generator tool~\cite{habets2006room}, with randomly chosen acoustic conditions,  such as room dimensions, microphone and speaker positions, and reverberation level. The parameters controlling the acoustic conditions can be found in Table~\ref{table:reverb_parameters}.  The noise signals were drawn from the WHAM! corpus \cite{wichern2019wham}, which consist of babble noises from different locations (such as restaurants, caf\'{e}s, bars, and parks) and added to the clean mixtures with random \ac{SNR} in the range of [10,25]~dB. Note that these acoustic conditions represent low reverberation conditions and relatively high \ac{SNR}.


\subsection{Algorithm Settings}
The speech and noise signals are downsampled to 8~[KHz].  The frame-size of the \ac{STFT} is  256 samples with $75\%$ overlap.  Due to the symmetry of the \ac{DFT} only the first half of the frequency bands is used.  $\beta_{\text{SISDR}}$ was set to  $0.75$ to emphasize the \ac{SI-SDR} loss. 

In the training procedure, we used the Adam optimizer~\cite{kingma2014adam}. The learning rate was set to 0.001 and the training batch size to 16. The weights are randomly initialized, and the lengths of the signals were randomly changed at each batch.


\subsection{Evaluation Measures}
To evaluate the proposed algorithm we use three evaluation metrics: 1) the \ac{SI-SDR}, as mentioned above, 2) the \ac{SIR}, and 3) the \ac{SDR}. These metrics are widely used for \ac{BSS} tasks. Note that for the \ac{SI-SDR} measure we present the improvement  (SI-SDRi). The proposed algorithm is compared to the commonly used VoiceFilter \cite{wang2018VoiceFilter} algorithm and to a recently proposed \ac{BSS} method \cite{chazan2021single} that demonstrates high performance even in reverberant and noisy conditions.  Additionally, we tested a variant of the proposed method in which, rather than the \ac{RI} features, only the log-spectrum is estimated while the noisy phase is used. Finally, an oracle solution was generated by using the target log-spectrogram and the noisy phase. This oracle solution provides the best achievable performance by the masking-based procedure. 

\subsection{Results}
\begin{figure}[h]
\centering
\includegraphics[trim=25 10 20 50 ,clip, width=0.45\textwidth,height=0.18\textheight]{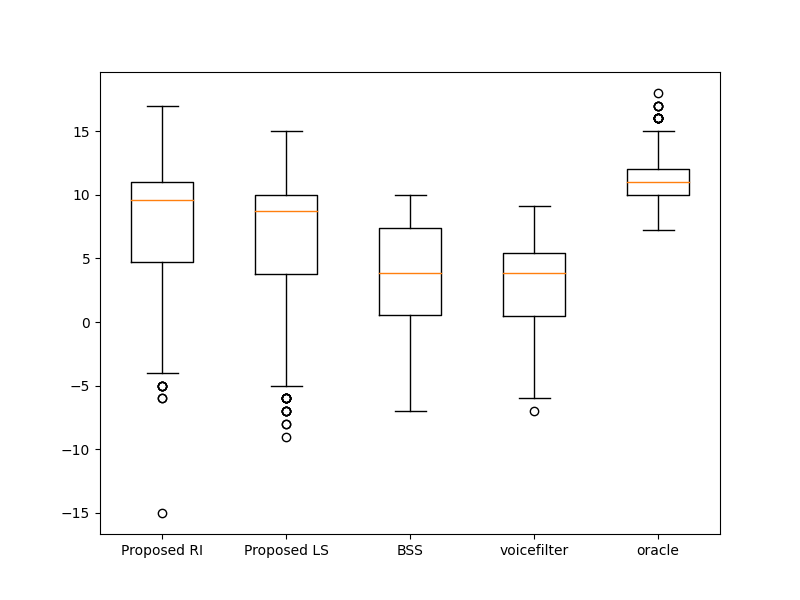}
    \caption{SI-SDRi comparison between the models for the clean test dataset. Two variants of the proposed method, (real-imaginary (RI) and log-spectrogram (LS) features) are compared to  VoiceFilter and to the Oracle masking-based method.}
    \label{fig:sisdri}
\end{figure}

\noindent\textbf{Clean conditions:}
The SI-SDRi results for the clean test dataset are depicted in Fig.~\ref{fig:sisdri}. First, it is easy to verify that the reference signal indeed assists the extraction task. Second, it is clear that the proposed approach outperforms the VoiceFilter algorithm. Finally, it can be deduced that the \ac{RI} features are more suitable to the task at hand than the log-spectrum (LS) features.
%
The \ac{SIR} and \ac{SDR} values are presented in Table \ref{table:bss_eval}.  It is evident that the proposed algorithm (RI variant) and the algorithm in \cite{chazan2021single} perform similarly in terms of \ac{SIR} measures. However, in the \ac{SDR} measure, the proposed algorithms clearly outperform the algorithm in \cite{chazan2021single}, implying lower distortion.

\begin{table}[htbp]
\caption{SIR and SDR results (the higher the better). }
\begin{center}
\begin{tabular}{@{}ccccccc@{}}
\toprule
  Model & Mixture &  BSS \cite{chazan2021single}  &  Proposed (LS)  & Proposed (RI)  \\
  \midrule
SIR   & 0.1 & \textbf{15.5} &  14.7 & 15.4 \\ 
SDR   &0.1 &  5.73 & 7.9 & \textbf{8.22} \\
\bottomrule
\end{tabular}
\end{center}
\label{table:bss_eval}
\end{table}


\begin{figure}[h]
\center
\subfigure[ $x(l,k)$]{\includegraphics[trim=0 10 0 30 , clip, width=0.48\textwidth,height=0.17\textheight]{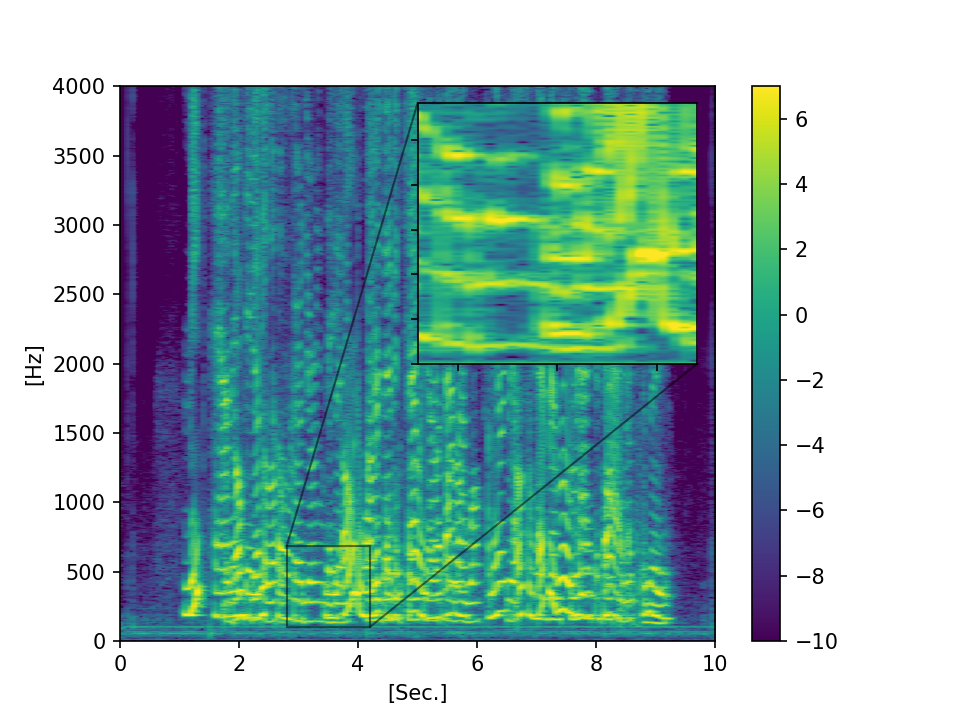}}\\
\subfigure[$\hat{s}_1(l,k)$]{\includegraphics[trim=0 10 0 30 , clip, width=0.48\textwidth,height=0.17\textheight]{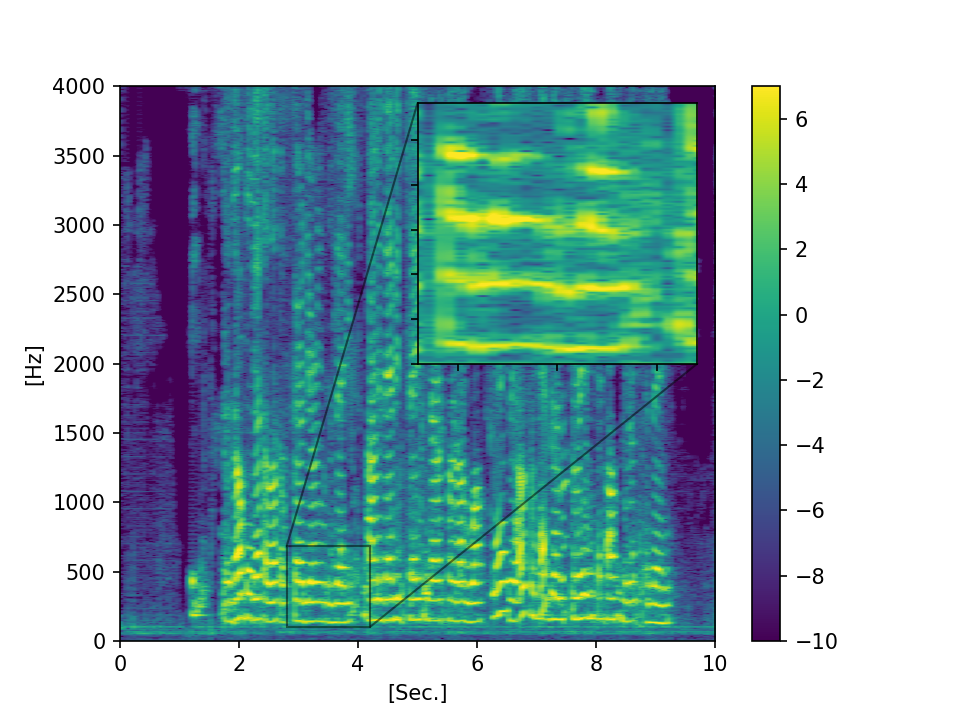}}\\
\subfigure[$\hat{s}_2(l,k)$]{\includegraphics[trim=0 10 0 30 , clip, width=0.48\textwidth,height=0.17\textheight]{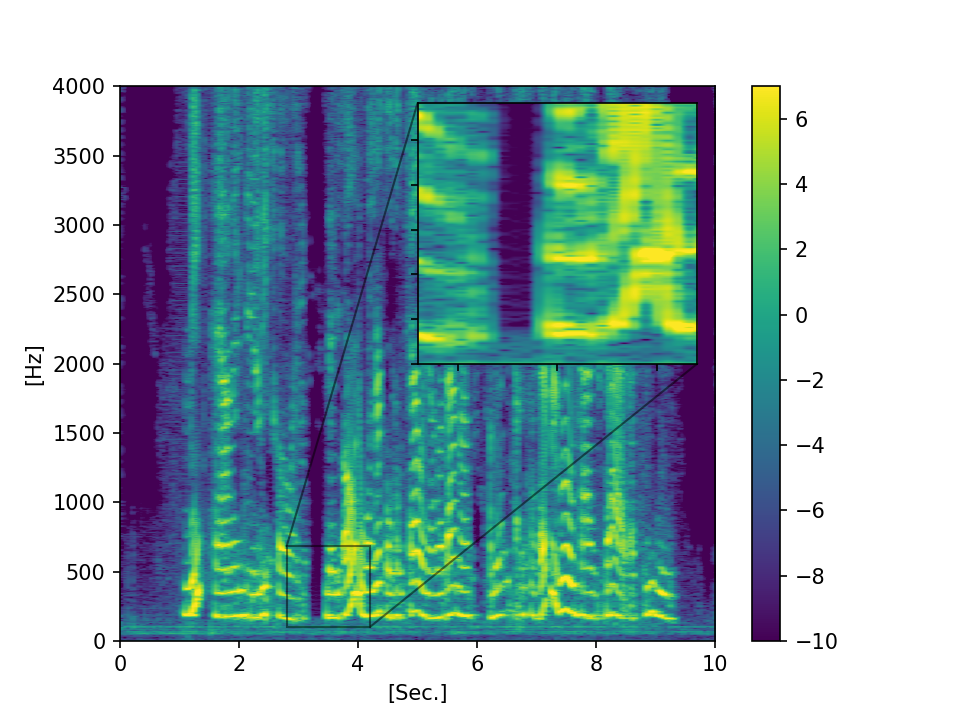}}
\caption{Real mixture recording and the extraction of  each speaker given its own reference signal.}
\vspace{-20pt}
    \label{fig:sonograms}
\end{figure}
\noindent\textbf{Noisy and reverberant conditions:}
Table~\ref{table:noisy_results} depicts the results of the proposed algorithm in comparison with the \ac{BSS} algorithm trained on noisy and reverberant conditions, and to the oracle log-spectrogram combined with the noisy phase. First, it is clear that the noisy phase with the oracle spectrogram deteriorates the extraction capabilities. It is worth noting that this is the maximum score that can be obtained by the masking-based approach. For this reason the performance of the VoiceFilter is not reported in this section. Second, our method outperforms the \ac{BSS} method.

\begin{table}[t]
\caption{Noisy reverberant data specification.}
\label{table:noisy_data}
\centering
\resizebox{0.8\columnwidth}{!}{
\begin{tabular}{@{}lll@{}}
\toprule
                        & $L_x$         & $U[4,8]$                                 \\
Room dim.~[m]                & $L_y$        & $U[4,8]$                                  \\
                        & $L_z$        & $U[2.5,3]$                                               \\ \midrule
Reverb. time~[sec]             &     $T_{60}$      & $U[0.16,2]$                            \\ \midrule
                        & $x$        & $\frac{L_x}{2}+{U}[-0.5,0.5]$ \\
Mic. Pos. [m]           &$y$         & $\frac{L_y}{2}+{U}[-0.5,0.5]$ \\
                        & $z$        & 1.5                                         \\ \midrule

Sources Pos. [$^\circ$] & $\theta$ & \emph{U}{[}0,180{]}                                \\ \midrule
Sources Distance [m]                   &          & $1+{U}[-0.5,0.5]$                        
            \\ \bottomrule
\end{tabular}}
\label{table:reverb_parameters}
\end{table}

\begin{table}[htbp]
\caption{SI-SDRi for noisy-reverberant data. }
\begin{center}
\begin{tabular}{@{}ccccccc@{}}
\toprule
  Model &  \ac{BSS} & Oracle &  Proposed \\
\midrule
Value   & 4.9 & 3.7 & \textbf{5.4}  \\
\bottomrule
\end{tabular}
\end{center}
\label{table:noisy_results}
\end{table}

\noindent\textbf{Real recording}
To further examine the capabilities of the proposed method, we recorded 2 speakers in a $3\times3\times2.5$, relatively quiet, enclosure. Both speakers are standing close to the microphone while uttering English sentences. Additionally, each participant was separately recorded to be used as the reference signal.  Fig.~\ref{fig:sonograms} depicts the sonograms of the experiment. The upper figure depicts the mixture recordings. The middle figure depicts the output of the model with the first reference. It is clear that the model accurately extracts the first speaker (denoted $\hat{s}_1$). To better understand the role of the reference signal embedding, the reference of the second speaker was recorded in Hebrew, which has different phonemes structure than in English. The lower figure demonstrates the extraction capabilities of the second speaker (denoted $\hat{s}_2$). It is easy to verify that the algorithm is still capable of extracting this speech signal, despite the use of a reference signal in a different language. This may imply that the embedding focuses on the speaker's characteristics rather than the content of the reference. The results are available for listening in our website.\footnote{https://sharongannot.group/audio/}

\section{Conclusions}
A novel combined time and time-frequency model was presented. This architecture enables the exploitation of the \ac{TF} patterns of the speech while utilizing the time-domain \ac{SI-SDR} loss. We also show that the \ac{RI} features are beneficial for clean and for noisy and reverberant conditions {and achieve better results than the LS features, which use the noisy phase for reconstruction of the wave signal}.  Experiments show that our model outperforms \ac{SOTA} \ac{BSS} algorithms as well as common speaker extraction models.

 \balance
\bibliographystyle{IEEEtran}
\bibliography{my_library}

\end{document}